\documentclass[11pt]{article}
\usepackage{setspace, graphics, feynmp, epsfig}
\newcommand{\be}{\begin{equation}}
\newcommand{\ee}{\end{equation}}
\newcommand{\bea}{\begin{eqnarray}}
\newcommand{\eea}{\end{eqnarray}}

\begin{document}

\begin{titlepage}
\def\thepage {}        

\title{The Standard Model hierarchy, fine-tuning, and negativity of the Higgs mass squared}

\author{
Marko B. Popovic\thanks{Other address: 
Artificial Intelligence Lab, Massachusetts Institute of Technology, Technology Square 200, Cambridge, MA  02139. E-mail: marko@ai.mit.edu}, \\
Department of Physics, Lyman Lab, room 428, Harvard University, \\
Cambridge, MA 02138. E-mail: marko@lorentz.harvard.edu}
\date{\today}

\maketitle
 
\bigskip
\begin{picture}(0,0)(0,0)
\put(295,250){HUTP-02/A012}
\put(295,235){hep-ph/0204345}
\end{picture}
\vspace{24pt}

\begin{abstract}

I discuss standard motivation for the new physics at the 1 TeV scale. Although the arguments for new exotic phenomena seem to be very supportive I argue that the Standard Model still might offer a good description far beyond this energy scale. I analyze three Standard Model cases with specific boundary conditions at the lowest energies. These conditions essentially eliminate the hierarchy and fine-tuning problems. In the first model where quartic scalar interactions are required to decouple the Higgs is predicted to weigh around $39$ GeV. In the second model with composite Higgs the top Yukawa coupling is required to be one (by hand, i.e. reflecting the assumed ground state) and the Higgs mass of about $138$ GeV is favored. The third model has condensation mechanism embedded in two dimensions. The top Yukawa coupling being one comes about as prediction rather then requirement, i.e. $g_t={3g_2 \over 2}\sqrt{1+{1\over3}\left(g_1\over g_2\right)^2}\;(1-\textit{few}\%)\approx 1.025\;(1-\textit{few}\%)$ where $g_2$, $g_1$ are electroweak $SU(2)\times U(1)$ gauge couplings, and the SM Higgs is expected to weigh in between $114.8$ and $118.6$ GeV.

\pagestyle{empty}
\end{abstract}
\end{titlepage}

\section{Introduction}
\setcounter{equation}{0}

The Standard Model (SM) of particle physics is in very good agreement with all experimental observations made thus far. Nevertheless, a common view continues that the SM can not be a valid, complete theory of Nature for energies just above the one TeV scale. 

Where does this view come from? Why does the model that so elegantly fits all low energy phenomena need to be embedded into a yet unknown but supposedly more fundamental framework for energies around the TeV scale? Traditionally, it is trusted that the reason lies in the SM hierarchy and fine-tuning problems. 

First I review the hierarchy and fine-tuning problems in their traditional form. 
Second I discuss the more realistic SM case. I do so by using two approaches which give identical results: the Veltman method with a hard Euclidean cutoff and a method that uses effective potential in the dimensional regularization minimal substraction ($\overline{MS}$) scheme. I explain why these approaches are equivalent and how they offer two different interpretations of the fine-tuning problem. 

Then, I suggest how the hierarchy and fine-tuning problems might be resolved by the SM itself, i.e. without a need for supplementing exotic new physics at the one TeV scale. The only new, however key, element is the presence of at least one additional requirement on the structure of the theory at the low energies. The trivial reason for the additional requirement is that it may institute the effective SM (including the Higgs mass) to be a self-contained, self-consistent structure up to highest energies. Finally, I analyze three models on these lines: Toy, Master and Interesting standard models. The favorite Higgs pole mass predictions are $39$ GeV, $138$ GeV, and about $116$ GeV respectively.

\section{The Hierarchy and Fine-Tuning: Warm Up With Traditional Views} 
 
The well-known non-zero vacuum expectation value (vev) of the scalar Higgs field, $v_{EW}=2.462 \times 10^2$ GeV, sets the scale of the electroweak interactions. Another well-known energy scale is the Planck scale, $\Lambda_{planck}\approx 1.2 \times 10^{19}$ GeV, probably setting the scale of the unified theory incorporating gravity. Obviously, many orders of magnitude separate these numbers. Traditionally, this fact is considered a serious obstacle for the SM. The problem termed hierarchy problem expresses doubts that the SM alone can provide a good physical description over such a broad range of energies.

Why is that a problem? It is well known that many physical theories give excellent descriptions of the phenomena over a broad range of distance and energy scales. The Newtonian theory of gravity covers more than fifteen orders of distance magnitude. With electromagnetism and photon propagation the range is even more impressive. So, why would the SM be the exception and fail at the energies in the reach of the next generation of accelerators? The traditional, incomplete answer is that the Higgs (along with Z and W's) field is massive and the photon is massless. In other words, the scalar mass squared receives an additive renormalization contribution; mass squared is not protected from the quadratic ``infinities!" 

But so what? This fact itself does not make the existing hierarchy look any worse or any better. The truth is that hierarchy does not constitute a problem of any kind. Only when united with the fine-tuning problem can the hierarchy look persuasive as an obstacle to the SM at high energies. 

Chronologically, the SM fine-tuning problem was introduced when technicolor models originated \cite{Susskind}. The logarithmic running of the strong technicolor gauge coupling is offered as a solution to this problem. The problem in its original form is reproduced now.

To characterize the fine-tuning problem it is useful to introduce the dimensionless mass parameter $\mu=m_H^2(\Lambda)/\Lambda^2$. The parameter $m_H(\Lambda)$ represents a renormalized Higgs mass at the cut-off energy scale $\Lambda$. The mass running is rudely simplified by   
\be
m_{H0}^2-m_H^2=g^2(\Lambda_0^2-\Lambda^2) \,\,\, ,
\ee
where $\Lambda_0$ is some large fundamental scale at which new physics (i.e. a more fundamental framework) parameterizes the SM. Mass runs quadratically, $g$ is approximated with a constant (!), and non-zero vev is formed when $m_H^2\approx \lambda v_{EW}^2$, i.e. $\mu_{EW}\approx \lambda$; here, in the same spirit, scalar's quartic coupling $\lambda$ is approximated with a constant (!).

Now, one may solve for $\mu_0=\mu(\Lambda_0)$ with $h=\Lambda_0^2/v_{EW}^2$ and obtain 
\be
\mu_0=g^2 + (\lambda-g^2)h^{-1} \,\,\, .
\ee
Clearly, if a large hierarchy, for example at the Planck scale where $h=\Lambda^2_{planck}/v^2_{EW}\sim 10^{34}$, is plugged into this formula a tremendous ``fine-tuning" is found. In other words, new physics at the scale $\Lambda_0=\Lambda_{planck}$ could ``potentially" position $\mu_0$ anywhere in between $g^2$ and $\lambda$ (if outside this region, the electroweak symmetry would be unbroken), but new physics has chosen a value that is tuned to $g^2$ with a precision of one in $10^{34}$. In addition, the more the theory is finely tuned the larger the hierarchy becomes. This is roughly sketched in Fig. 1.a. 

If there is no reasonable explanation (by the SM alone or new physics) that may accommodate such tremendous tuning (and with such small numbers one would think that there is none) the conclusion is that $h$ should not be very large. Therefore, from this line of reasoning, i.e. fine-tuning as originally postulated, it may be concluded that the SM can be a good description only up to very small energies, maybe ten or so times larger than a physical Higgs mass!
\begin{figure}
\centerline{\epsfig{file=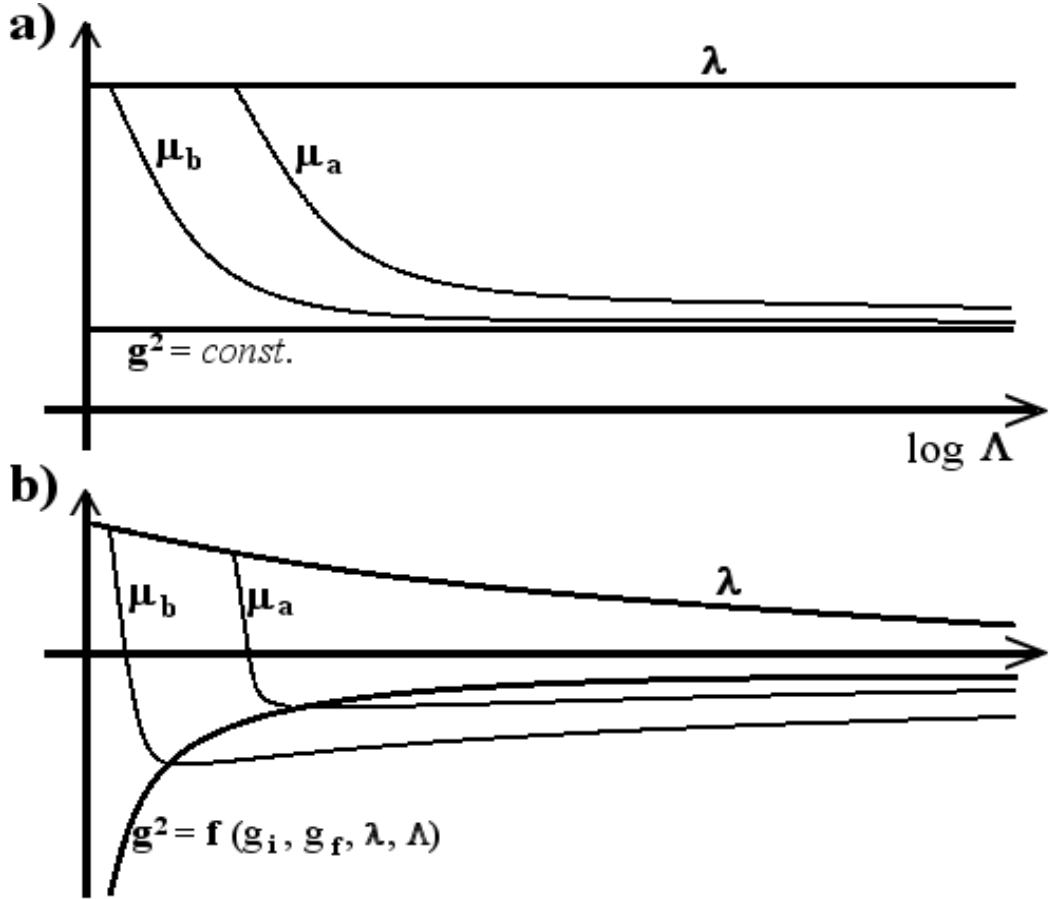, height=12cm,width=14cm,angle=0}}
\caption[lepton]{Illustration of the fine-tuning problem in traditional sense (top sketch) and in the more realistic SM case (bottom sketch). The purpose of these drawings is to show the qualitative distinctions between the functional forms of the two cases; by no means should the two plots be quantitatively compared or separately analyzed. The gauge and Yukawa couplings are labeled with $g_i$ and $g_f$ respectively while $\mu_a$ and $\mu_b$ identify the two ``potential" evolutions of the dimensionless mass parameter.}
\label{Moose}
\end{figure}

\section{The Hierarchy and Fine-Tuning: The Standard Model Realm}

The quantity $m_H(\Lambda)$, introduced above, is sometimes described as a quantity that as a matter of principle \cite{LiCoPhy} cannot be calculated. This suggests that the whole fine-tuning problem cannot be quantified (ill-posed problem?). A matter of principle refers to the fact that the calculation of $m^2_H(\Lambda)$ is regularization scheme dependent; if one chooses two different and supposedly equally good regularization schemes, one might get two different and supposedly equally good answers. 

The solution to this problem is to pick one regularization scheme and then check whether a tremendous fine-tuning exists at all. Clearly, this is a legitimate thing to do - the tremendous fine-tunning should not be regularization scheme dependent! Here we will use two methods and show that the results are essentially identical.


In the hard cutoff Euclidean regularization scheme the integrals of the type
\be
\int{d^d p} \; , \;\; \int{d^d p \over p^2} \;\; \ldots
\ee
are nonzero while in dimensional regularization due to the dilatation property are identically zero. In logarithmic terms the two regularization methods agree. This interplay may be easily seen if one dimensionally continues and regularizes propagators by the method of Pauli-Villars. Then one finds \cite{ZinnJustin} for example for $d<4$
\be
I_{\gamma}=\int{{d^dq \over (2\pi)^2}{1 \over {q^2(p+q)^2}}}\sim{\left[(p^2)^{(d/2)-2}-\Lambda^{d-4}\right] \over {8\pi^2(4-d)}}
\ee
Fixing the cutoff and taking the limit $d=4$ one obtains $\ln{\Lambda}$ instead of a pole at $d-4$. Vice versa, by fixing $d-4$ and taking the cutoff to infinity one obtains the continuation of the initial integral.

As is standard, in the hard Euclidean cutoff method (which does not support the Lorentz symmetry) one obtains the running scalar mass that runs quadratically. In contrast, dimensional regularization results in only logarithmic running. One may think that this property of dimensional regularization may solve the fine-tuning problem. However, that is not the case. They are really just two different ways of organizing terms in the expansion or if you prefer, two different interpretations.

The fine-tuning may be analyzed from the $\overline{MS}$ dimensional regularization point of view. The quantities $m_H^2$ and $f=dm_H^2(\Lambda)/d\Lambda^2$ (paralleling the ``constant" $g^2$ in the previous section) may be calculated using the $\overline{MS}$ scheme effective potential $V_{eff}$ at the one-loop level, with all coupling constants running logarithmically at the two-loop level. Here, it is important to make a distinction between the $\overline{MS}$ parameters $m$ and $m_H$. The $\overline{MS}$ mass parameter $m$, as is standard, has intrinsic logarithmic running. The parameter $m^2_H$ running quadratically is analyzed in the context of the fine-tuning problem. It is defined by the tree-level form of the effective potential, i.e. $V_{eff}(\phi_{cl})=V(\phi_R)=-m^2_H(\Lambda \sim \phi_R) \phi^2_R /4 +\lambda(\Lambda \sim \phi_R) \phi^4_R /8$, where $\phi_{cl}$ and $\phi_R$ stand for classical and renormalized scalar fields.
So to speak $m_H^2$ is a quantity where one loop correction to the effective potential is lumped together with the $\overline{MS}$ parameters $m$. Evolution of the SM dimensionless mass parameter $\mu$ for the physical Higgs masses in the range $140-230$ GeV is shown in the Fig. 2..

\begin{figure}
\centerline{\epsfig{file=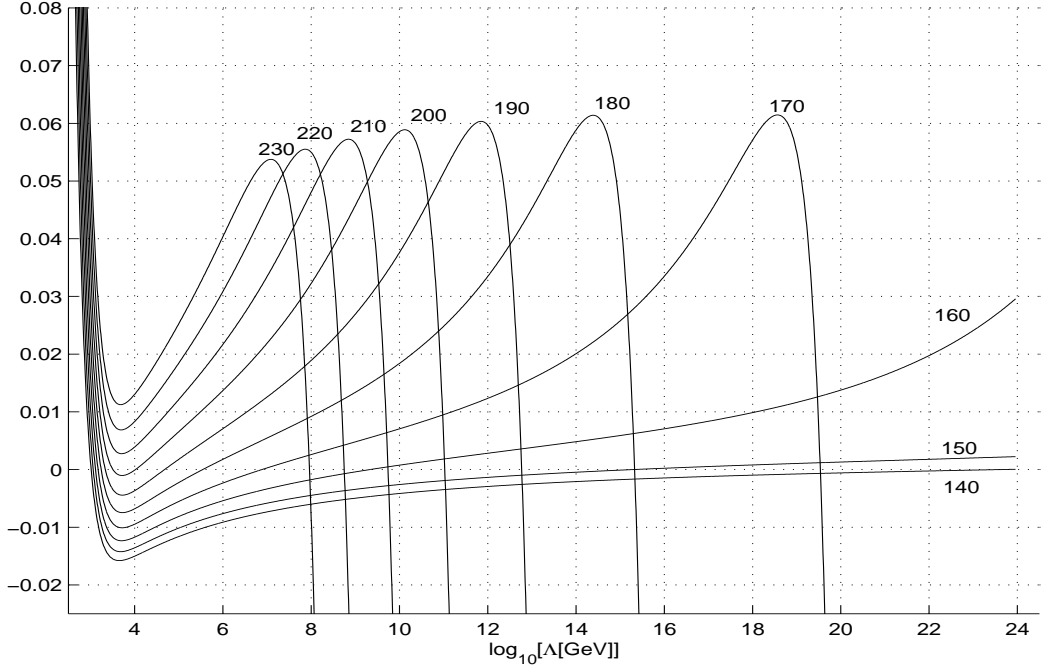, height=9cm,width=14cm,angle=0}}
\caption[lepton]{Evolution of the SM dimensionless mass parameter $\mu$ for the physical Higgs masses in the range 140 - 230 GeV. The regions around and beyond maxima should be considered unphysical due to the perturbativity constraint \cite{Popovic}.}
\label{Moose}
\end{figure}

Although both methods give the same result (with exceptional agreement) it is probably more instructive to look at the fine tuning problem from the perspective of the hard Euclidean cutoff method. As done in \cite{Veltman} one could express the quadratic running as polynomial in the SM masses 
\be
{dm_H^2 \over d\Lambda^2}={{3g_2^2} \over {64\pi^2M_W^2}}\left[m_H^2+2M_W^2+M_Z^2-4\sum\left(n_f \over 3\right) m_f^2\right] +\ldots \; . \label{Higgs1}
\ee
where $n_f=3(1)$ for quarks(leptons). The same expression may be rewritten as polynomial in the SM couplings (where the tree level relationship $\lambda=m_H^2/v_{EW}^2$ is used) 
\be
{dm_H^2 \over d\Lambda^2}=-{1 \over {32 \pi^2}}\left[ 12 g_t^2 -6\lambda-{9 \over 2}g_2^2-{3 \over 2} g_1^2 \right]+\ldots \; . \label{Higgs2}
\ee
The quantities $\mu$ and $f=dm_H^2(\Lambda)/d\Lambda^2$ are by definition related as
\be
{d\mu \over {d \ln (\Lambda^2)}}=f-\mu \,\,\, . \label{master}
\ee
Obviously, if $f$ is a constant it would represent an ultraviolet fixed point. However $f$ is scale dependent and this introduces some small difference. For large $\Lambda$ the quantity $\mu$ quickly approaches the function
\be 
f_{UV\;fixed}=f-f^{(1)}+f^{(2)}-\ldots
\ee
where all higher derivatives of $f$, i.e. $f^{(n)}=d^nf/d(\ln \Lambda^2)^n$, are also expressible as polynomials in the SM couplings\footnote{Strictly speaking only the one loop level running has a clear physical meaning (i.e. the couplings may be entangled only at the one loop level \cite{Pirogov}). The higher loops are affected by the regularization method - however these corrections, in sense of the final result are rather negligible. However, we do assume that there is a preferred way how Nature is expressing physically observable effects at high energies.}. Therefore, given the SM couplings we may imagine that we could calculate $ f_{UV\;fixed}$ and then if we look at the physics of very large $\Lambda$, we should realize that the parameter $\mu$ is extremely close to the value of this function at that scale. We may also answer how close by calculating the difference 
\be
\mu(m_{H(pole)}^2)- f_{UV\;fixed}=\lambda(m_{H(pole)})(1+\textit{few}\%) \; .
\ee
The size of this difference is important because it tell us the particular solution $\mu_p=\textit{const.}/\Lambda^2$ of the differential equation (\ref{master}). Combining these two equations we get
\be
\textit{const.}\approx m_{H(pole)}^2
\ee
Therefore at some larger scale $\Lambda$ we may write 
\be
\mu(\Lambda^2)= f_{UV\;fixed}(\Lambda^2)+{ m_{H(pole)}^2 \over \Lambda^2} \; .
\ee
If we vary obtained solution by some function $\delta \mu$ we also see that 
\be
{d\delta\mu \over {d \ln (\Lambda^2)}}=-\delta\mu \,\,\, .
\ee
In other words tiny perturbation at large scale causes gigantic changes at low energies.

At which scale is the theory said to be finely tuned?
One may define the measure of the fine tuning via 
\be
\mu_p(\Lambda^2) /(\lambda(\Lambda^2)- f_{UV\;fixed}(\Lambda^2)) \;\; .
\ee 
Both quantities in the denominator run slowly, i.e. logarithmically, and therefore the one percent tuning is reached quickly, an order of magnitude or so above the $m_{H(pole)}$.

However, why is the size of the particular solution so important? The point is that it is not defined by the SM couplings (running logarithmically) and therefore the new physics should explain it somehow. The question is how the new physics can explain the quantity that is getting so small at the largest scales. 
Well isn't that a good sign that the SM should be ruled out at energies above the TeV scale if we want things to look natural? Isn't that a good sign that the new physics scale must be one TeV?

Well, not necessarily. The case may be that we asked a wrong question. It should be clear by now that we don't really worry about the existence of fixed point behavior but really about the size of the particular solution and its connection to the new physics. Maybe the SM itself may explain the size of particular solution - then no new physics is needed to perform that job and fine-tuning is removed. Clearly the new physics could exist but its scale then doesn't need to be one TeV. I discuss the above possibility in the next section and analyze three models where this property is implemented.

\section{Toy, Master and Interesting Standard Models}

The SM structure seems to be too perfect and quite general to fail at some finite energy scale-it clearly supports (and is built on) the idea of the infinite range of validity. In a sense the fine-tuning problem itself originates from this fact. If we believe that something new should happen at least around the Planck scale, we run the parameters of the theory and then we see that the theory is finely tuned there. This is a consequence of the SM structure and the presence of the cutoff that we introduced. Maybe what we need is a less perfect and a less general four dimensional theory that is somehow more compatible with the cutoff idea and/or we need an additional constraint on the theory. 
\\
\\
{\bf \textit{a) Toy Standard Model}}
\\
\\
Consider first a toy model which in every aspect resembles the SM except for the missing $\Phi^4$ interactions at the minimum of the effective potential. On the classical tree level this theory obviously cannot exhibit the phenomenon of electroweak symmetry breaking. However, on the quantum level it can. This is possible if there is, for example, one fermion with large Yukawa coupling - translated to the SM numerology, i.e. using observed $W$ and $Z$ masses, this situation corresponds to the existence of a fermion with a mass larger than roughly $\sqrt{3/n_c} \; 72.6$ GeV (where $n_c$ is the number of colors), i.e. Yukawa coupling that is larger than $\sqrt{3/n_c} \; 0.41 \approx \sqrt{3/n_c} \; \sqrt{1/6}$. Well, that is promising - we have exactly such a candidate at our disposal. The above numerology follows from the Higgs mass running, equations (\ref{Higgs1}-\ref{Higgs2}), \footnote{It is interesting to note that if both space time dimension and Dirac trace were set to 2 (they were both set to 4 in the original calculation \cite{Veltman}) the Higgs mass running would be proportional to $6m_t^2-2M_W^2-M_Z^2-3m_H^2$. If that is the right thing to consider, then the above result changes and the minimal fermion mass needs to be $\sqrt{3/n_c} \; 59.3$ GeV, or in terms of the Yukawa coupling $\sqrt{3/n_c} \; 0.34$.} and the minimum of the effective potential which in this case is roughly given as
\be
{dV_{eff} \over \Phi}=-{1 \over 2}{m_H^2}\Phi-{1 \over 2}{dm_H^2 \over \Lambda^2}|_{\Lambda\sim\Phi}\Phi^3 \; . 
\ee

What does this funny theory have to do with the fine-tuning problem? Basically, it explains why the hierarchy can be so large. In the effective constrained SM picture of this type the electroweak symmetry is by definition driven by the ``technical" parameter $\lambda$ running logarithmically that at some scale, which we then call vev, goes to zero (more precisely to $-(1/4)d\lambda/\ln\Lambda$ so that quartic sector is removed). Or alternatively, by neglecting the quartic contributions altogether, by the scale at which one Yukawa coupling becomes large enough that it overcomes the effect of gauge bosons!  

It is amusing to note that this toy model gives a prediction of the Higgs mass that is in very good agreement with global electroweak precision fit prediction coming from ``all data" except hadronic asymmetry determination of the effective leptonic weak mixing angle (sets C and D in \cite{Chanowitz2}), i.e. $m_H \approx 39.2$ GeV. This value is obtained from the second derivative of $V_{eff}$ (with $d\lambda/\ln\Lambda$ term being considered as well). 
\\
\\
{\bf \textit{b) Master Standard Model}}
\\
\\
Another example is the nonrenormalizable model where the fundamental Higgs sector is essentially excluded. Here the non-SM interactions are expected to introduce a specific type of dimensional transmutation through color singlet SM fermion attractive channels that trigger electroweak symmetry breaking at low energies. Without specifying the type of these interactions we will assume this situation to be describable as effective SM theory of a composite scalar field which gets the vev in the usual way through the interplay of effective $\mu$ and $\lambda$ parameters. Again, as in the last example there will be an additional constraint on this structure at the electroweak symmetry breaking scale (more precisely at the scale of electroweak vev) - we will require the sum of Yukawa couplings squared/top Yukawa coupling to be approximately one (up to smallish radiative corrections). Well this sounds promising as we have exactly such a situation in hand!
 
Instead of bothering with the notorious complication of many flavors, we will concentrate on just the top sector with $T_L=(t, b)_L$ and $T_R=t_R$ and consider a color singlet composite scalar field that has exactly the right quantum number of the usual Higgs scalar field. This scenario share some features of the top-mode model of \cite{Bardeen}. However, instead of expecting parameters of this model to essentially blow up at the high energy compositeness scale, as in \cite{Bardeen}, we will require perturbative, smooth decoupling of the scalar sector at high energies, i.e. $\mu,\lambda \rightarrow 0$ in actual calculation. In addition we set by definition (or if you prefer by hand) the top Yukawa couplings to be one. By doing so we assume that $g_t\approx1$ is not accidentally related to the ultraviolet loop dynamics but rather that it represents an important characteristic of the vacuum.

What does this situation have to do with the fine-tuning problem? It removes it! The fact that electroweak symmetry breaking is now defined by two constraints means that scalar mass is completely defined at any energy scale by the SM itself. It cannot be varied as two constraints ($g_t=1$ and $V_{eff}^(1)=0$) would not coincide. This picture means that our type of universe belongs to a much smaller subset of universes defined by the general SM structure (with a requirement on the minimum of the effective potential alone) - this fact indeed is what can make Planck mass hierarchy natural. 


Our main accent here is clearly to present the possibility of the existence of an additional defining principle for our effective SM realization that solves the hierarchy and fine-tuning problems. We are not trying to give the physical description of potential non standard physics at high energies and its exact matching to the effective SM theory in the infrared - that would require much work for even a decent understanding. However we could make a small excursion for reasons of completeness and hypothesize that as in the top-mode model \cite{Bardeen} one can start at high energies with the ``true fundamental" or again effective theory (reflecting however the ground state) containing dimension six 4-fermion operators like
\be
{\cal{L}}_{int}\sim \left(\bar{T}T\right) \left(\bar{T}T\right)
\ee
and no fundamental scalar field.\footnote{At this point it is worth noticing that this is a renormalizable interaction in two dimensions.} One may then use the old recipe \cite{Eguchi} and rewrite this interaction with the use of static auxiliary scalar fields and obtains \textit{static} scalar mass squared terms plus Yukawa terms. That is in principle a valid procedure if the classical equations of motion are indeed satisfied at the cutoff scale (and therefore one would expect the one loop corrections to be \textit{small} and physical description to be \textit{perturbative}). Going toward smaller energy scales, one then obtains the non-zero quartic scalar coupling terms and dynamical scalar fields. It is believed that this theory is strongly attracted to the effective SM infrared description. 

Concentrating only on the top sector, these scalar fields will be represented by nine weak doublet fields; nine due to the color combinations and weak doublets because of the $\overline{T_L}T_R$ form. The eight color sensitive fields decouple quickly below the cutoff due to the strong interaction (and are represented as a tremendously massive object in the low energy description, i.e. without breaking the color interactions)\footnote{At the cutoff scale their fixed point is somewhat larger, due to the strong interactions, than the one for a color invariant scalar field. Therefore they necessarily gain huge \textit{negative} mass squared and decouple an order or so below the cutoff. For them the standard logic is applicable - they are not \textit{naturally} finely tuned at the cutoff scale.} while a color invariant scalar, i.e. standard Higgs,
\be
\Phi={1 \over \sqrt{3}}\left(\Phi_{red-red}+\Phi_{blue-blue}+\Phi_{green-green}\right)
\ee
survives and triggers the electroweak symmetry breaking. The contribution to the Z, W vacuum polarization sets $v_{color-color}=v_{ew}/\sqrt{3}\approx 142$ GeV. 

How about the leftover degree of freedom corresponding to the propagating physical Higgs particle? What is the mass of the Higgs in this picture? The natural guess would be that it should be of the order $v_{ew}/\sqrt{3}\approx 142$ GeV. It is interesting to realize that if the condition on vacuum stability, in particular on quadratic divergences, is imposed on each color separately\footnote{Possibility that quadratic divergences cancels if sensitivity on color interaction is omitted was considered in the past \cite{Al-sarhi,Ma}.} one obtains the relationship 
\be
{4 \over 3} m_t^2-2M_W^2-M_Z^2-m_H^2=0 \label{vacuum}
\ee    
implying the Higgs mass $m_H \approx 138.9$ GeV!\footnote{It is hard not to notice that this value happens to be extremely close to $v_{ew}/\sqrt{\pi}$. However this is probably just a numerical curiosity.}  The contribution from top loops  changes by factor $(n_f/3) (1/\sqrt{3})^2=1/3$ compared to equation (\ref{Higgs1}). The electroweak gauge bosons contributions is unchanged due to identical electroweak coupling. The quartic contribution is also unchanged as may be directly checked. 

What is missing here is the standard picture on the composite objects, i.e. bound states. What is Higgs exactly made of? How could it have a mass smaller than the top quark? What is top made of? We believe that the answer might be related to space time dimensionality of the field theory needed for the appropriate description of the electroweak symmetry breaking. Anyhow, we analyze next the SM renormalization flow and try to infer proper answers from the very high energies.

Naively one can simply require the quartic coupling to go to zero and in this way define the cutoff scale versus physical Higgs mass. This has some chance to be a good description but one needs to be aware that close to the cutoff scale the SM renormalizion flow could be a bit deceptive - partially due to the cutoff effects by themselves and also due to the presence of the color octet fields and maybe some other objects that we are not aware of at this point. Until this cutoff sensitivity is properly physically understood the results obtained from pure SM flow must be taken with a grain of salt. Anyhow, the SM solution incorporating quartic scalar coupling that goes to zero at some scale below the Planck scale exists for the physical Higgs masses roughly smaller than 138 GeV! The larger the physical Higgs mass the larger the cutoff scale. That is a well-known result for the quartic coupling flow and we refer interested reader to references \cite{Popovic}. In this picture any Higgs pole mass roughly below 138 GeV is allowed and the heavier Higgses are excluded. 

However, some spice can be added to this picture by requiring in addition that quantum corrections need to be smaller than the tree-level values as we approach 
the cutoff energy. From the perspective of $\overline{MS}$ scheme that would suggest that 
\begin{figure}
\centerline{\epsfig{file=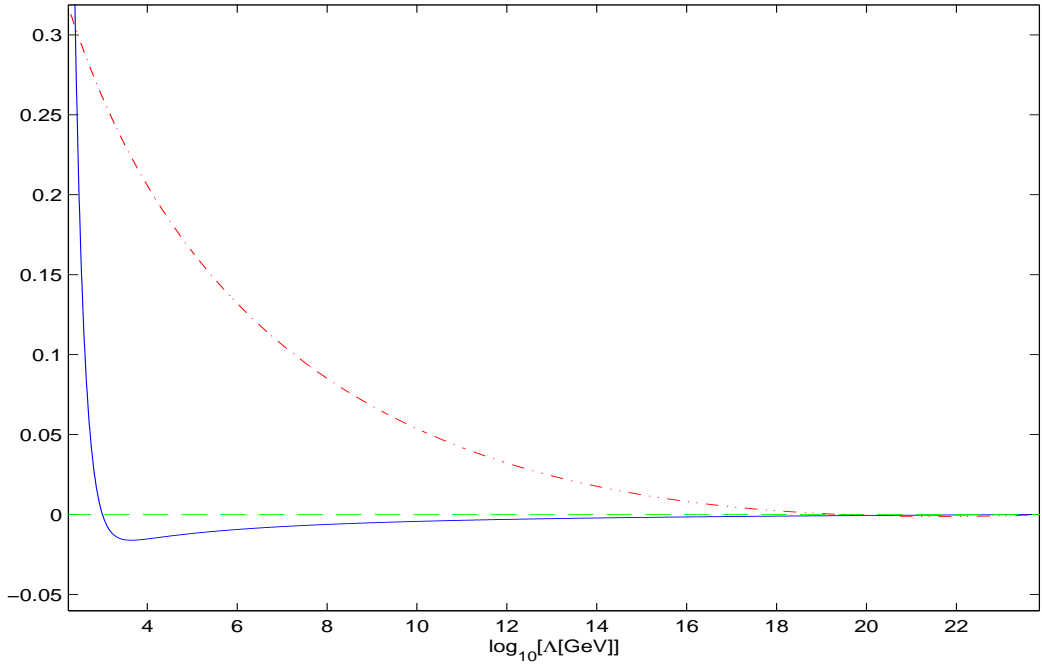, height=9cm,width=14cm,angle=0}}
\caption[lepton]{The SM dimensionless mass parameters $\mu$ (solid line) and quartic scalar coupling $\lambda$ (dashed-dotted line) characterizing the effective scalar potential for the physical Higgs mass $m_H=137.6$ GeV. Dashed line represents zero value.}
\label{Moose}
\end{figure}
one loop correction, i.e. parameter $\mu \Lambda^2$ roughly, needs to be on the order of the running $\overline{MS}$ mass at the cutoff scale. It is worth pointing out that the presence of the small parameter $\mu$ has nothing to do with an explanation of the fine-tuning - without requiring two constraints that define the electroweak breaking as a master principle, one is not able to explain why $\mu$ should be so close to $f_{UVfixed}$.

Numerically, to our level of precision, we notice that there is no solution that satisfies both conditions $\mu=0$ and $\lambda=0$ at any cutoff scale.\footnote{The parameter $\mu$ at the Planck scale is equal zero for $m_H=146$ GeV.} The manner in which the curves (do not) intersect reminds one somewhat of the GUT case \cite{Georgi}. As shown in Fig. 3. the three curves $f_1=\mu$, $f_2=\lambda$ and $f_3=0$ come extremely close in the vicinity of the Planck scale being the cutoff and for the physical Higgs masses in the vicinity of 138 GeV. As the author does not know of a better experimentally motivated cutoff scale it is the author's strong feeling that this observation constitutes a good candidate for the mass of the last missing ingredient in the SM! 

As promised a few thoughts on the negativity of the Higgs mass squared are in order. The negativity of the effective Higgs mass should not be considered a bad feature of the theory. Nor should the theory be expected to necessarily break down at the energy scale at which there is a transition from positive to negative values. It is not the pole mass that is negative and the real physical description does not suffer from the tachionic degrees of freedom. It is simply a mass squared in the effective propagator of the effective tree-level description defined at a particular energy scale. Nothing more and nothing else. If effective mass is negative the process of interest will be more suppressed at that energy scale. 

It is particularly useful to consider this ``negativity" from the unbroken phase point of view. Consider raising the temperature of the field theory. Through the standard Higgs mechanism this process is supposed to bring us to the true minimum of the $\Phi^4$ potential that lies at zero. And the natural description contains the broken phase negative mass squared that we now label as positive in an unbroken phase point of view. At energy scales much larger than the temperature, the mass squared is the same in both unbroken and broken theory descriptions (up to the minus sign). Therefore, if we associate some fundamental meaning to the non-zero temperature description and expect this theory to have an effective scalar mass squared of the same sign up to the cutoff scale then it should come with no surprise that broken phase contains negative effective mass squared at high energies as some type of memory on the unbroken phase physics. Again, we are not proving that this must be true. However, we find this to be  very natural feature of the healthy effective SM theory.
\\
\\
{\bf \textit{c) Interesting Standard Model}}
\\
\\
If the dynamical Higgs mechanism is fully realized in two dimensions, it opens the opportunity that the vacuum energy problem may be resolved - the condition on non-zero vev of the effective scalar field may be then confined only along the path of physically propagating objects. Here we take both space time dimension and Dirac trace to be two and reproduce the previous analysis for the Master SM (with now \textit{renormalizable} four fermion interactions).

The requirement on the vacuum stability, paralleling the one in equation (\ref{vacuum}), is now 
\be
2m_t^2-2M_W^2-M_Z^2-3m_H^2=0
\ee 
suggesting $m_H=114.8$ GeV. 

Here we conjecture the particular kind of equivalence in between the two pictures: toy model in two dimension with critical (\textit{two dimensional}) fermion Yukawa coupling and standard electroweak symmetry breaking in four dimensions. The hope is that the critical fermion may be thought as a constituent of the physical Higgs degree of freedom as standard relativistic composite algebra seems to suggest, i.e.
\be
m_H=2m_{critical}(1-\textit{few}\%) \; ,
\ee
where $m_{critical}=\sqrt{3/n_c} \; 59.3$ GeV.
Furthermore this opens a possibility that the top quark may be thought as a composite object, i.e. bound state of three two-dimensional critical fermions. As interactions are rather weak the natural guess on the top quark mass is then
\be
m_t=3m_{critical}(1-\textit{few}\%) \sim v_{EW}/\sqrt{2}
\ee
Finally the requirement on the quartic scalar coupling being zero suggests the cutoff scale of roughly $10^6$ GeV.

The picture with both Higgs and top quark being composite objects might have interesting implications. If we think on the scalar sector as an ``order parameter" tool we expect $<\Phi>=0$ in the unbroken phase (unbroken phase clearly dominates majority of the ``empty" space in this picture). The electroweak symmetry is triggered by the large Yukawa coupling in two dimensions that at some energy scale becomes critical - fermion contribution balances the gauge boson contributions to the $<\Phi>$. Immediately when fermion contribution start dominating i.e. $<\Phi>$ differs just infinitely small from zero, the process of condensation quickly develops and top quark and physical Higgs as bound states of massive fermions are formed! If that is the right picture then the correct way of thinking is not to impose the tree level $g_t=1$ as requirement but rather
\be
g_t \approx 3 \sqrt{{{9\over2}g_2^2+{3 \over 2}g_1^2} \over 18}\;(1-\textit{few}\%)={3g_2 \over 2}\sqrt{1+{1\over3}\left(g_1\over g_2\right)^2}\;(1-\textit{few}\%)\approx 1.025\;(1-\textit{few}\%)
\ee   
which is after all maybe just by accident so close to one. Anyhow, this relationship is our second constraint on the SM defining $v_{EW}$ (and $m_H$) and removing the hierarchy and fine-tuning problems.

Whether the two-dimensional large Yukawa coupling fermion, building the vacuum structure of the theory (expected not to suffer from the vacuum energy problem), can be observed as physically propagating object or not seems at this point very moot. From dimensional analysis we would conclude that this is not the case. However we postpone answering this question until better understanding is reached.

\section{Conclusion}

The SM hierarchy and fine-tuning problems may supplement valid motivation for the existence of drastically new physics description beyond the one TeV scale. However, the SM structure can also be a good candidate for a self-consistent high energy representation of Nature if additional relationship(s) at low energies is(are) realized. Then new physics is not needed to explain the size of the Higgs mass as this mass may be explained by the SM itself. If that is the case the new physics may become important only at very high energies.  

I presented three models on these lines. In the first exemplary toy model the scalar quartic interactions were required to decouple at the scale of the electroweak breaking vev so that the ground state is completely determined by the quadratic term and quantum radiative corrections. The physical Higgs mass in this toy model is predicted to be about $39$ GeV.

The second model suggests that there is no fundamental scalar but that the Higgs is a composite object composed out of the top quark degree of freedom. The top Yukawa coupling is required to be one at the lowest energies. The SM structure is run up to high energies at the two loop level and it is found that the requirement of the scalar sector completely decoupling is very closely reached for the Higgs mass in the vicinity of 138 GeV and the physical cutoff on the order of the Planck scale. The $m_H \approx 138$ GeV value roughly fall in the $90\%$ CL intervals from both leptonic and hadronic asymmetry measurements and it is in very good agreement with $\chi^2$ distribution obtained from $\Gamma_Z$ alone \cite{Chanowitz2}.

The third model has a two dimensional mechanism of condensation. Both Higgs and top quark are thought to be the composite objects. The top Yukawa coupling is predicted to be $g_t={3g_2 \over 2}\sqrt{1+{1\over3}\left(g_1\over g_2\right)^2}\;(1-\textit{few}\%)\approx 1.025\;(1-\textit{few}\%)$ where $g_2$, $g_1$ are electroweak $SU(2)\times U(1)$ gauge couplings and the Higgs mass is expected to be somewhat larger than $114.8$ GeV in an unpleasant, hopefully accidental, proximity to the late LEP Higgs signal candidate \cite{Analysis}.

If Higgs happens to weight either around $116$ or $138$ GeV, and if nature has chosen the large SM high-energy desert in front of us, the hopes for discovery of new phenomena should not be lost. It is my strong belief that this exciting challenge will motivate a new set of interesting, clever and successful ways to study nature. It might be a somewhat longer journey than the one with new physics waiting for us at the 1 TeV scale, but we may be surprised. 

\section*{ Acknowledgments }
\indent
I thank the particle physics theory groups at Harvard University and Boston University for their reactions to some of the ideas presented in this work. I am grateful to the biomechatronics group of MIT's AI Lab for partial financial support. I am thankful to Ms. Wendy Gu and Mr. Matthew Malchano for the proofreading of the manuscript. I am indebted to my family for their love.

\end{document}